\input harvmac
\input epsf
%
%
\noblackbox

\newcount\figno
\figno=0
\def\fig#1#2#3{
\par\begingroup\parindent=0pt\leftskip=1cm\rightskip=1cm\parindent=0pt
\baselineskip=11pt
\global\advance\figno by 1
\midinsert
\epsfxsize=#3
\centerline{\epsfbox{#2}}
\vskip -21pt
{\bf Fig.\ \the\figno: } #1\par
\endinsert\endgroup\par
}
\def\figlabel#1{\xdef#1{\the\figno}}
\def\encadremath#1{\vbox{\hrule\hbox{\vrule\kern8pt\vbox{\kern8pt
\hbox{$\displaystyle #1$}\kern8pt}
\kern8pt\vrule}\hrule}}

\def\frac#1#2{{#1 \over #2}}

\def\p{\partial}
\def\semi{\subset\kern-1em\times\;}

\def\sqr#1#2{{\vcenter{\vbox{\hrule height.#2pt
\hbox{\vrule width.#2pt height#1pt \kern#1pt \vrule width.#2pt}
\hrule height.#2pt}}}}

\def\CD{{\cal D}}

\def\CO{{\cal O}}                   
                   
                   \def\CW{{\cal W}}

\def\p{\partial}

\def\At{\tilde{A}}
\def\Ot{\tilde{O}}

%

%
\def\a{\alpha}
\def\p{\partial}

\def\Dt{\tilde{D}}
\def\Ft{\tilde{F}}
\def\cht{\tilde{\chi}}
\def\Pt{\tilde{P}}
\def\L{\Lambda}
\Title{\vbox{\baselineskip12pt
\hbox{hep-th/0303104}
\hbox{UCLA-03-TEP-07}
\vskip-.5in}}
{\vbox{\centerline{On the Matter of the Dijkgraaf--Vafa Conjecture}}}
\centerline{Per Kraus and  Masaki Shigemori}
\bigskip\medskip
\centerline{\it Department of Physics and Astronomy,}
\centerline{\it  UCLA, Los Angeles, CA 90095-1547,}
\centerline{\tt pkraus, shige@physics.ucla.edu}
\medskip
\baselineskip18pt
\medskip\bigskip\medskip\bigskip\medskip
\baselineskip14pt
%

\noindent With the aim of extending the gauge theory -- matrix
model connection to more general matter representations, we prove that
for various two-index tensors of
the classical gauge groups,
the perturbative contributions to the glueball superpotential 
reduce to matrix integrals.   Contributing diagrams consist of
certain combinations of spheres, disks, and projective planes,
which we evaluate to four and five loop order.  In the case of
$Sp(N)$ with antisymmetric matter, independent results are
obtained by computing the nonperturbative superpotential for
$N=4,6$ and $8$. Comparison with the Dijkgraaf-Vafa approach
reveals agreement up to $N/2$ loops in matrix model perturbation
theory, with disagreement setting in at $h=N/2+1$ loops, $h$ being
the dual Coxeter number. At this order, the glueball superfield
$S$ begins to obey nontrivial relations due to its underlying
structure as a product of fermionic superfields.  We therefore find
a relatively simple example of an ${\cal N}=1$ gauge theory admitting a
large $N$ expansion, whose dynamically generated superpotential
differs from the one obtained in the matrix model approach.

\Date{March, 2003}

\lref\DijkgraafFC{
R.~Dijkgraaf and C.~Vafa,
``Matrix models, topological strings, and supersymmetric gauge theories,''
Nucl.\ Phys.\ B {\bf 644}, 3 (2002)
[arXiv:hep-th/0206255].
}

\lref\DijkgraafVW{
R.~Dijkgraaf and C.~Vafa,
``On geometry and matrix models,''
Nucl.\ Phys.\ B {\bf 644}, 21 (2002)
[arXiv:hep-th/0207106].
}

\lref\DijkgraafDH{
R.~Dijkgraaf and C.~Vafa,
``A perturbative window into non-perturbative physics,''
arXiv:hep-th/0208048.
}

\lref\DijkgraafXD{
R.~Dijkgraaf, M.~T.~Grisaru, C.~S.~Lam, C.~Vafa and D.~Zanon,
``Perturbative computation of glueball superpotentials,''
arXiv:hep-th/0211017.
}

\lref\ItaKX{
H.~Ita, H.~Nieder and Y.~Oz,
``Perturbative computation of glueball superpotentials for SO(N)
and  USp(N),''
JHEP {\bf 0301}, 018 (2003)
[arXiv:hep-th/0211261].
}

\lref\CachazoRY{
F.~Cachazo, M.~R.~Douglas, N.~Seiberg and E.~Witten,
``Chiral rings and anomalies in supersymmetric gauge theory,''
JHEP {\bf 0212}, 071 (2002)
[arXiv:hep-th/0211170].
}

\lref\WittenYE{
E.~Witten,
``Chiral Ring Of Sp(N) And SO(N) Supersymmetric Gauge Theory
In Four Dimensions,''
arXiv:hep-th/0302194.
}

\lref\ChoBI{
P.~L.~Cho and P.~Kraus,
``Symplectic SUSY gauge theories with antisymmetric matter,''
Phys.\ Rev.\ D {\bf 54}, 7640 (1996)
[arXiv:hep-th/9607200].
C.~Csaki, W.~Skiba and M.~Schmaltz,
``Exact results and duality for Sp(2N) SUSY gauge theories with an  antisymmetric tensor,''
Nucl.\ Phys.\ B {\bf 487}, 128 (1997)
[arXiv:hep-th/9607210].
}

\lref\VenezianoAH{
G.~Veneziano and S.~Yankielowicz,
``An Effective Lagrangian For The Pure N=1 Supersymmetric Yang-Mills Theory,''
Phys.\ Lett.\ B {\bf 113}, 231 (1982).
}

\lref\SeibergJQ{
N.~Seiberg,
``Adding fundamental matter to 'Chiral rings and anomalies in
supersymmetric gauge theory',''
JHEP {\bf 0301}, 061 (2003)
[arXiv:hep-th/0212225].  See references therein for previous work on
fundamental matter.
}

\lref\BalasubramanianTM{
V.~Balasubramanian, J.~de Boer, B.~Feng, Y.~H.~He, M.~x.~Huang, V.~Jejjala and A.~Naqvi,
``Multi-trace superpotentials vs. Matrix models,''
arXiv:hep-th/0212082.
}

\lref\ArgurioHK{
R.~Argurio, V.~L.~Campos, G.~Ferretti and R.~Heise,
``Baryonic corrections to superpotentials from perturbation theory,''
Phys.\ Lett.\ B {\bf 553}, 332 (2003)
[arXiv:hep-th/0211249];
I.~Bena, R.~Roiban and R.~Tatar,
``Baryons, boundaries and matrix models,''
arXiv:hep-th/0211271;
}

\lref\AshokBI{
S.~K.~Ashok, R.~Corrado, N.~Halmagyi, K.~D.~Kennaway and C.~Romelsberger,
``Unoriented strings, loop equations, and N = 1 superpotentials from  matrix models,''
arXiv:hep-th/0211291.
}

\lref\JanikNZ{
R.~A.~Janik and N.~A.~Obers,
``SO(N) superpotential, Seiberg-Witten curves and loop equations,''
Phys.\ Lett.\ B {\bf 553}, 309 (2003)
[arXiv:hep-th/0212069].
}

\lref\BrezinSV{
E.~Brezin, C.~Itzykson, G.~Parisi and J.~B.~Zuber,
``Planar Diagrams,''
Commun.\ Math.\ Phys.\  {\bf 59}, 35 (1978).
}

\lref\KlemmCY{
A.~Klemm, K.~Landsteiner, C.~I.~Lazaroiu and I.~Runkel,
``Constructing gauge theory geometries from matrix models,''
arXiv:hep-th/0303032.
}

\lref\DoreyTJ{
N.~Dorey, T.~J.~Hollowood, S.~Prem Kumar and A.~Sinkovics,
``Exact superpotentials from matrix models,''
JHEP {\bf 0211}, 039 (2002)
[arXiv:hep-th/0209089].
}

\lref\IntriligatorUK{
K.~A.~Intriligator,
Phys.\ Lett.\ B {\bf 336}, 409 (1994)
[arXiv:hep-th/9407106].
}


\newsec{Introduction}

The methods of Dijkgraaf and Vafa
\refs{\DijkgraafFC,\DijkgraafVW,\DijkgraafDH} represent a
potentially powerful approach to  obtaining  nonperturbative
results in a wide class of supersymmetric gauge theories.    Their
original conjecture consists of two parts. First, that holomorphic
physics is captured by an effective superpotential for a glueball
superfield, with nonperturbative effects included via the
Veneziano-Yankielowicz superpotential \VenezianoAH. Second, that
the Feynman diagrams contributing to  the perturbative part of the
glueball superpotential reduce to matrix model diagrams.

The second part of the conjecture has been proven for a few choices of
matter fields and gauge groups, namely $U(N)$ with adjoint
\refs{\DijkgraafXD,\CachazoRY} and
fundamental \SeibergJQ\ matter,
and $SO/Sp(N)$ with adjoint matter \refs{\ItaKX,\AshokBI,\JanikNZ}.
Combining this with
the first part of the conjecture has then been shown to reproduce known gauge
theory results.  Some examples of ``exotic'' tree-level superpotentials have
also been considered successfully, such as multiple trace \BalasubramanianTM\
and baryonic \ArgurioHK\ interactions.

One naturally wonders how far this can be pushed.
Generic ${\cal N}=1$ theories possess intricate dynamically
generated superpotentials which are difficult or (nearly) impossible to
obtain by traditional means, and so a systematic method for computing
them would be most welcome.  The promise of the DV approach is that these
perhaps can be obtained to any desired order by evaluating matrix integrals.
With this in mind, we will demonstrate the reduction to matrix integrals
for some new matter representations. We will then find some impressive
agreements, as well as obstacles, 
 when comparing to known gauge theory results


In particular, it is straightforward to generalize the results of
\refs{\DijkgraafXD,\ItaKX} to more general two-index tensors of
$U(N)$ and $SO/Sp(N)$, with or without tracelessness conditions imposed.
The relevant $0+0$ dimensional Feynman diagrams which one needs to compute
consist of various spheres, disks and projective planes, and
disconnected sums of these.  We evaluate these to five-loop order.

For comparison with gauge theory we focus on the particular case
of $Sp(N)$\foot{Our convention for $Sp(N)$ is such that
$N$ is an even integer, and
$Sp(2) \approx SU(2)$.}
with an antisymmetric tensor chiral superfield.
The dynamically generated superpotentials for  such theories are
highly nontrivial, and cannot be obtained via the ``integrating in''
approach of \IntriligatorUK.   Furthermore, the results display no
simple pattern in $N$.  Nevertheless, a method is known for computing
these superpotentials on a case-by-case basis \ChoBI.   Results for
$Sp(4)$ and $Sp(6)$ were obtained in \ChoBI, and here we extend this
to $Sp(8)$ as well (partial results for $Sp(8)$ appear in \ChoBI).  
We believe that these examples illustrate the
main features of generic ${\cal N}=1$ superpotentials, and so are
a good testing ground for the DV approach.

For our $Sp(N)$ examples, we will demonstrate agreement between our
gauge theory superpotentials and the DV approach up to 
$N/2$ loops in perturbation
theory, with a disagreement setting in at $N/2+1$ loops.  In terms
of the glueball superpotential, we thus find a disagreement at
order $S^h$, where $h=N/2+1$ is the dual Coxeter number of
$Sp(N)$.

Given that discrepancies occur, it is perhaps not surprising that they arise
at order$S^h$,
for it is at this order that $S$ begins to obey relations due to
its being a product of two fermionic superfields \refs{\CachazoRY,\WittenYE}.
Furthermore, at
this order contributions to the effective action for $W_\a$ of the
schematic form ${\rm Tr~} (W_\a)^{2h}$ can be reexpressed in terms of
lower traces, including $S^h$.  Unfortunately, it is not clear how to
ascertain these relations {\it a priori}, since they receive corrections from 
nonperturbative effects (see \WittenYE\ for a recent discussion).
These complications do not arise for theories with purely adjoint matter,
since unlike in our examples, the gauge theory  results are known to have  
a simple pattern in $N$, and so $N$
can be  formally taken to infinity to avoid having to deal with any
relations involving the $S$'s.  
There are also a number of other potential subtleties involved, as we  will
discuss in section 5. In any case, it seems
that additional input is required to make progress at $h$ loops and beyond.

The remainder of this paper is organized as follows.  In section 2 we
isolate the  field theory diagrams that contribute to the
glueball superpotential, derive the reduction of these 
diagrams to those of a
matrix model,  and discuss their computation.   These
results are used in section 3 to derive effective superpotentials
for $Sp(N)$ with matter in  the antisymmetric tensor representation.
In section 4 we state the corresponding results derived from a
nonperturbative superpotential for these theories.  Comparison
reveals a discrepancy, which we discuss in section 5.  Appendix A
gives more details on diagram calculations; appendix B collects
results from matrix model perturbation theory; and appendix C
concerns the computation of dynamically generated superpotentials
for the $Sp(N)$ theories.

%

\newsec{Reduction to matrix model}

In this section we will extend the results of
\refs{\DijkgraafXD,\ItaKX} to include the following matter
representations: \vskip.15cm

\noindent $\bullet$~ $U(N)$ adjoint.

\noindent $\bullet$~ $SU(N)$ adjoint.

\noindent $\bullet$~  $SO(N)$ antisymmetric tensor

\noindent $\bullet$~  $SO(N)$ symmetric tensor, traceless or
traceful.

\noindent $\bullet$~ $Sp(N)$ symmetric tensor.

\noindent $\bullet$~  $Sp(N)$ antisymmetric tensor, traceless or
traceful.

\vskip.15cm

 We will use $\Phi_{ij}$ to denote the matter
superfield.    In the case of $Sp(N)$, $\Phi_{ij}$ is defined as
\eqn\bh{ \Phi=\cases{ SJ & $S_{ij}$: symmetric tensor,\cr AJ &
$A_{ij}$: antisymmetric tensor. } }
Here $J$ is the invariant antisymmetric tensor of $Sp(N)$, namely
\eqn\bha{ J_{ij}= \pmatrix{0 & {\bf 1}_{N/2}\cr -{\bf 1}_{N/2} &
0}. }
The tracelessness of the $Sp$ antisymmetric tensor is defined
with respect to this $J$, i.e., by $\Tr[AJ]=0$.

The fact that allows us to treat the above cases in parallel to
those considered in \refs{\DijkgraafXD,\ItaKX} is that gauge
transformations act by commutation,  $\delta_\Lambda \Phi \sim
[\Lambda, \Phi]$.  A separate analysis is needed for, say, $U(N)$
with  (anti)symmetric matter (see \KlemmCY\ for some work on such cases). 

\subsec{Basic Setup}

Following \DijkgraafXD, we consider a supersymmetric gauge theory
with chiral superfield $\Phi$ and field strength $\CW^\a$.
Treating $\CW^\a$ as a fixed background, we integrate out $\Phi$
to all orders in perturbation theory.  We are interested in the
part of the effective action which takes the form of a
superpotential for the glueball superfield $S={1\over 32\pi^2} \Tr
[\CW^\alpha \CW_\alpha]$. In \DijkgraafXD, using the superspace
formalism, it was shown that this  can be obtained from a simple
action involving only chiral superfields:
\eqn\bd{ S(\Phi)= \int \!d^4p\, d^2\pi\,\left[
\frac{1}{2}\Phi(p^2+\CW^\alpha \pi_\alpha)\Phi +W_{\rm tree}(\Phi)
\right]. }
 We choose the tree level superpotential to be
\eqn\bha{ W_{\rm tree}=\frac{m}{2}\Tr[\Phi^2]+{\rm interactions},
}
where the interactions are single trace terms,  and include the
mass in the propagator:
\eqn\bfa{
\frac{1}{p^2+m+\CW^\alpha \pi_\alpha}~.
}

Actually, we have to be a little more precise here.  Displaying
all indices, we can write the quadratic action as
\eqn\bfb{
\frac{1}{2}\int d^4p \, d^2\pi\, \Phi_{ji} G^{-1}_{ijkl} \Phi_{kl}
}
with
\eqn\bfc{
G^{-1}_{ijkl}=\left[(p^2+m)\delta_{im}\delta_{jn}+(\CW^\alpha)_{ijmn}\pi_\alpha
\right]P_{mnkl}.}
Here the $P$'s are projection operators appropriate for the gauge
group and  matter representation under consideration:
\eqn\bfd{ P_{ijkl}= \cases{ \delta_{ik}\delta_{jl} & $U(N)$
adjoint \cr  & \cr \delta_{ik}\delta_{jl}- {1 \over
N}\delta_{ij}\delta_{kl} & $SU(N)$ adjoint \cr & \cr {1\over
2}(\delta_{ik}\delta_{jl} - \delta_{il}\delta_{jk})
  & $SO(N)$ antisymmetric \cr & \cr {1\over
2}(\delta_{ik}\delta_{jl} + \delta_{il}\delta_{jk})
  & $SO(N)$ traceful symmetric \cr & \cr {1\over 2} (\delta_{ik}\delta_{jl} + \delta_{il}\delta_{jk} - {2 \over
N}\delta_{ij}\delta_{kl})
  & $SO(N)$  traceless symmetric \cr & \cr
{1\over 2}(\delta_{ik}\delta_{jl} - J_{il}J_{jk})
  & $Sp(N)$ symmetric \cr & \cr
{1\over 2}(\delta_{ik}\delta_{jl} + J_{il}J_{jk})
  & $Sp(N)$ traceful antisymmetric\cr & \cr
{1\over 2}\left(\delta_{ik}\delta_{jl} + J_{il}J_{jk} -
\frac{2}{N}\delta_{ij}\delta_{kl}\right)
  & $Sp(N)$ traceless antisymmetric}}
The propagator is then given by the inverse of $G^{-1}$ in the
subspace spanned by $P$:
\eqn\bfe{\langle \Phi_{ji} \Phi_{kl}\rangle = \left[ { P \over p^2
+m + \CW^\a \pi_\a}\right]_{ijkl}= \left[ \int_0^\infty \! ds ~
e^{-s(p^2+m + \CW^\a \pi_\a)} P\right]_{ijkl}~. }
Our rule for multiplying four-index objects is
$(AB)_{ijkl}=\sum_{mn} A_{ijmn} B_{mnkl}$.  The fact that gauge
transformations act by commutation means that we can write
\eqn\bff{ (\CW^\a)_{ijkl} = (\CW^\a)_{ik}\delta_{jl} -
(\CW^\a)_{lj}\delta_{ik}}
where on the right hand side $(\CW^\a)_{ij}$ are field strengths
in the defining representation of the gauge group.

\subsec{Diagrammatics}

The presence of the three sorts of terms in the projection
operators \bfd\ means that in double line notation we have three
types of propagators, displayed in Fig. 1.
\vskip-.3cm
%
%
\fig{Propagators.  a) untwisted;  b) twisted; c) disconnected}
{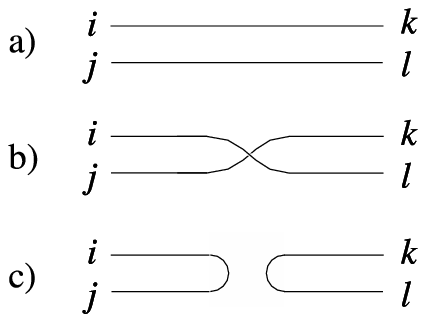}{5cm}
\figlabel{\diagfig}
  Note in
particular the disconnected propagator, which allows us to draw
Feynman diagrams which have disconnected components in index space
(All diagrams are connected in momentum space since we are
computing the free energy).  A typical diagram involving cubic
interactions is shown in Fig. 2.
\vskip-.9cm
%
%
\fig{Typical diagram}{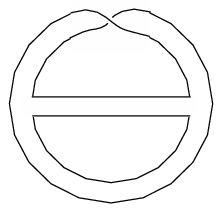}{3.5cm}
\figlabel{\diagfig}
Since we are computing
the superpotential for $S$, we include either zero or two
insertions of $\CW^\a$ on each index loop.\foot{Note that we are
explicitly {\it not} including the contributions coming from more
than two $\CW^\a$'s on an index loop, even if for a particular $N$
these can be expressed in terms of $S$'s.  We will come back to
this point in section 5.} We will now prove that the
diagrams which contribute are those consisting of some number of
sphere, disk, and projective plane components.  Furthermore, the
total number of disconnected components must be one greater than
the number of disconnected propagators.   Fig. 3 is an
example of a contributing diagram,
\vskip-.4cm
%
%
\fig{Contributing diagram}{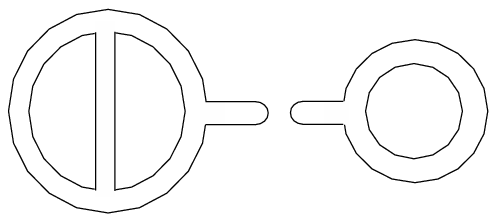}{5cm}
\figlabel{\diagfig}
\noindent
while Fig. 4 is a
diagram which does not contribute.
%
%
\fig{Non-contributing diagram}{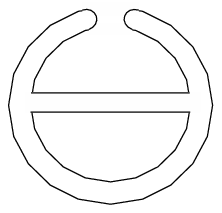}{3.5cm}
\figlabel{\diagfig}

The proof is is similar to that given in
\refs{\DijkgraafXD,\ItaKX}, so we mainly focus on the effect of
the new disconnected propagator. In double line notation we
associate each Feynman diagram to a two-dimensional surface.  Let
$F$ by the number of faces (index loops); $P$ be the number of
edges; and $V$ be the number of vertices.  The Feynman diagram
also has some number $L$ of momentum loops.
 Euler's theorem
tells us that
\eqn\bb{F= P- V + \chi}
where $\chi_{S^2} = 2$, $\chi_{D^2}=\chi_{RP^2} = 1$.  We also
have the relation
\eqn\bba{ F =L -1 +\chi.}
In a  diagram with $L$ loops we need to bring down $L$ powers of
$S$ to saturate the fermion integrals, and we allow at most one
$S$ per index loop.   Therefore, for a graph to be nonvanishing we
need $F \geq L$.    Graphs on $S^2$ with no disconnected
propagators have $F= L+1$, and those on $RP^2$ have $F=L$.

To proceed we will make use of the following operation.
Considering some diagram $D$ that includes some number of
disconnected propagators.  To each $D$ we associate a diagram
$\tilde{D}$, obtained by replacing each disconnected propagator of
$D$ by an  untwisted propagator.  Each $\Dt$ diagram thus consists
of a single connected component.  $D$ and $\Dt$ have the same
values of  $L$ and $V$, but can have different values of $F$, $P$,
and $\chi$.  We use $\Ft$, $\Pt$,  and $\cht$ to denote the number
of faces, edges, and the Euler number of $\Dt$.  To see which
diagrams can contribute we consider various cases. \vskip.1cm

\noindent {\bf Case 1:}  $D$ has no disconnected propagators, so
$\Dt=D$.  This case reduces to that of \refs{\DijkgraafXD,\ItaKX},
and so we know that only $S^2$ and $RP^2$ graphs contribute (since
no $D^2$ graphs can arise without disconnected propagators, these
are the only graphs for which $F\geq L$.)

\vskip.1cm

The remaining cases to consider are those for which we have at
least one disconnected propagator.

\vskip.1cm

\noindent {\bf Case 2:} $\chi =\cht \leq 1 $

\vskip.1cm In this case $\Ft \leq L$, from \bba. Each time we take
a disconnected  propagator and replace it by an untwisted
propagator we are increasing $P$ by $1$  but keeping $L$
unchanged. Therefore, from \bb, this operation increases $F$ by
$1$. So we see that in this case $F < L$. This means that the
diagram $D$ does not contribute.

\vskip.1cm

\noindent {\bf Case 3:}  $\chi= \cht =2$

\vskip.1cm   $\Dt$ has $\Ft=L+1$.  In this case, if $D$  has a
single disconnected propagator, we will have $F=L$, and so the
diagram might seem to contribute. But we will now show that the
fermion determinant vanishes for such diagrams.

We follow the conventions of \ItaKX, where the reader is referred
for more details.    The fermion contribution is proportional to
$[\det N(s)]^2$ where
\eqn\bc{N(s)_{ma} = \sum_i s_i K^T_{mi}L_{ia}~.}
Here, $i$ labels propagators; $m$ labels ``active'' index loops on
which we insert an $S$; and $a$ labels momentum loops. In the
present case, since $F=L$, all index loops are active and so $N$
is a square matrix.   To show that the determinant vanishes, we
will show that the rectangular matrix
\eqn\bd{ s_i K_{im}}
has a nontrivial kernel.

 Recall the definition of $K_{im}$. For
each oriented propagator labelled by $i$, the $m$th index loop can
do one of three things: 1) coincide and be parallel, giving
$K_{im}=1$; 2) coincide and be anti-parallel, giving $K_{im}=-1$;
3) not coincide, giving $K_{im}=0$.    Consider $K_{im}$ acting on
the vector $b_m$ whose components are all equal to $1$. It should
be clear that
\eqn\be{\sum_m K_{im} b_m = 1-1=0~.}

The intuitive way to think about this is that $b_m$ are the index
loop momenta and $\sum_m K_{im}b_m$ are the propagator momenta. By
setting all index loop momenta equal, one  makes all propagator
momenta vanish, and this corresponds to an element of the kernel
of \bd.   This finally implies $\det N(s)=0$, which is what we
wanted to show.

\vskip.1cm

\noindent {\bf Case 4:}  $\chi \neq \cht$ \vskip.1cm

 This can only happen when  $D$ has two or more disconnected components.
     In this case, $\chi > \cht$ and so \bb\ still
allows $F \geq L$ even when $P<\Pt$.

In order to have a nonvanishing fermion integral, each component
of $D$ must have $F \geq L$, so each component must be an $S^2$, a
$D^2$, or an $RP^2$.   Suppose $D$ has $N_{S^2}$ $S^2$ components,
$N_{D^2}$ $D^2$ components, and $N_{RP^2}$ $RP^2$ components, so
that
\eqn\gb{\chi = 2N_{S^2}+ N_{RP^2} +N_{D^2}~.}

Next consider the relation between $P$ and $\Pt$. The number of
disconnected  propagators must be at least the number of
disconnected components of $D$ minus one, so
\eqn\gc{P = \Pt -(N_{S^2}+ N_{RP^2} + N_{D^2} -1) - a = \Pt +1
+N_{S^2} -\chi -a}
where $a$ is a nonnegative integer.  Now use
\eqn\gd{F  = P-V+\chi =\Pt +1 +N_{S^2}- V -a~.}
$\Dt$ satisfies
\eqn\ga{\Pt+1 -V = L}
so we get
\eqn\gee{ F = L+N_{S^2} -a~. }

Now, in order to have a nonzero fermion determinant we need to
have at least one inactive index loop (no $\CW^a$ insertions) per
$S^2$ component after choosing $L$ active index loops.  In other
words, a nonvanishing fermion determinant requires
\eqn\gf{F \geq L+N_{S^2}~.}
Putting these two conditions together, we clearly need $a=0$. This
says that the number of disconnected  propagators in $D$  must be
precisely equal to the number of disconnected components of $D$
minus one.

\noindent{}

\noindent {\bf Summary:}  Diagrams which contribute to the
glueball superpotential have any number of disconnected $S^2$,
$D^2$, and $RP^2$ components.  The number of disconnected
propagators must be one less than the number of disconnected
components.\foot{It is easy to convince oneself that disconnected
diagrams will therefore never contribute in theories with only
even powers in the tree level superpotential, thus giving the same 
glueball superpotential in the traceful and traceless cases.}

\subsec{Computation of Diagrams}

Now that we have isolated the class of diagrams which contribute
to the glueball superpotential, we turn to their computation. This
turns out to be a simple extension of what is already known.  In
particular, the contribution from a general disconnected diagram
is simply equal to an overall combinatorial factor times the
product of the contributions of the individual components.   This
follows from the fact that, for the diagrams we are considering,
the disconnected propagators carry vanishing momentum, so the
diagrams are actually disconnected in both momentum space and
index space.

Next, we observe that the stubs from the disconnected propagators
can be neglected in the computation; it is easily checked that the
sum over  $\CW^a$ insertions on the stubs  gives zero due to the
minus sign in \bff.

So we just need rules for treating each component individually,
and then we multiply the contributions together to get the total
diagram. The rules consist of relating the gauge theory
contribution to a corresponding matrix contribution.  The cases of
interest are:

\vskip.1cm

\noindent {\bf $S^2$ components:}~ From the work of \DijkgraafXD,
we know that if $N^{L+1} F_{S^2}^{(L)}(g_k)$ is the contribution
in the matrix model diagram from an $L$ loop $S^2$ graph, then the
contribution in the gauge theory diagram is
\eqn\gfa{ W_{S^2}^{(L)}(S,g_k) = (L+1)N S^L F_{S^2}^{(L)}(g_k)}
  The prefactor $(L+1)N$ comes from the choice of, and trace
over, a single  inactive index loop.

\vskip.1cm

\noindent {\bf $RP^2$ components:}~  From the work of \ItaKX, we
know that if $N F_{RP^2}^{(L)}(g_k)$ is the contribution in the
matrix model diagram from an $L$ loop $RP^2$ graph, then the
contribution in the gauge theory diagram is
\eqn\gfb{ W_{RP^2}^{(L)}(S,g_k) = \pm 4 S^L F_{RP^2}^{(L)}(g_k)}
 The prefactor of $\pm
4$ comes from the fermion determinant, and is equal to $+4(-4)$ for
symmetric(antisymmetric) tensors.

\vskip.1cm

\noindent {\bf $D^2$ components:}~ These have $L=0$ and hence no
$\CW^a$ insertions.  So if the contribution to the matrix model is
$N F_{D^2}^{(L)}(g_K)$ then
\eqn\gfc{W_{D^2}^{(L)}(g_k)=N F_{D^2}^{(L)}(g_K)}

With the above rules in hand, it is a simple matter to convert a
given matrix model Feynman diagram into a contribution to the
glueball superpotential.   The example given in Appendix A should
help to clarify this.  We should emphasize that the above
procedure must by done diagram by diagram --- there is no obvious
way to directly relate the entire glueball superpotential to the
matrix model free energy; the situation is similar to
\BalasubramanianTM\ in this respect.

\newsec{Results from Matrix Integrals}

The considerations thus far apply to any single trace, polynomial,
tree level superpotential.
We now restrict attention to cubic interactions,
\eqn\za{ W_{\rm tree}=\frac{m}{2}\Tr\,\Phi^2 +\frac{g}{3} \Tr\,\Phi^3.}
(which are of course trivial in the case of $SO/Sp$ with adjoint
matter.) In  Appendix B we collect our matrix model results for
the various matter representations.  In this section we focus on
two particular cases, which  will be compared to gauge theory
results in the next section.

\subsec{Sp(N) with traceful antisymmetric matter}

The perturbative part of the glueball superpotential for $Sp(N)$
with traceful antisymmetric matter is
\eqn\bm{\eqalign{
W_{\rm traceful}^{\rm pert}(S,\alpha)
&=
\left(-N+3\right)\alpha S^2
+\left(-{16\over 3}N+{59\over 3}\right)\alpha^2 S^3\cr
&+\left(-{140\over 3}N+197\right)\alpha^3 S^4
+\left(-{512}N+{4775\over 2}\right)\alpha^4 S^5
+\cdots}
}
with %
\eqn\bmbb{\alpha\equiv {g^2 \over 2 m^3}.}
 In terms of diagrams,
\bm\ represents the contribution from $2,3,4$ and $5$ loops.
According to the DV conjecture, the full glueball superpotential
is then $W^{\rm eff} = W^{VY}+W^{\rm pert}$, where $W^{VY}$ is the
Veneziano--Yankielowicz superpotential:
\eqn\bma{
W^{VY}=(N/2+1) S[1-\log(S/\Lambda^3)]~.
}
 We are now instructed to extremize $W^{\rm
eff}$with respect to $S$ and substitute back in.  We call the
result $W^{\rm DV}$. Working in a power series in $g$, we obtain
\eqn\bmb{\eqalign{
 W_{\rm traceful}^{\rm DV}(\Lambda,m,g)
 =&
 (N/2+1)\Lambda^3
 \Bigg[
 1
 -\frac{2(N-3)}{N+2}\Lambda^3\alpha
 -\frac{2(4N^2+45N-226)}{3(N+2)^2}\Lambda^6\alpha^2\cr
 &-\frac{2(12N^3+293N^2+368N-8340)}{3(N+2)^3}\Lambda^9\alpha^3\cr
 &-\frac{96N^4+3803N^3+25868N^2-85092N-744768}{3(N+2)^4}\Lambda^{12}\alpha^4
 -\cdots
 \Bigg]
}}
For $N=4,6,8$, this yields
\eqn\bn{\eqalign{
 W^{{\rm DV},Sp(4)}_{\rm traceful}(\Lambda,\alpha) &=
 3\Lambda^3
 -{}\Lambda^6\alpha
 -{}\Lambda^9\alpha^2
 +{353 \over 27}\Lambda^{12}\alpha^3
 +{25205 \over 81}\Lambda^{15}\alpha^4
 +\cdots
 \cr
 W^{{\rm DV},Sp(6)}_{\rm traceful}(\Lambda,\alpha) &=
 4\Lambda^3
 -{3}\Lambda^6\alpha
 -{47\over 6}\Lambda^9\alpha^2
 -{73 \over 2}\Lambda^{12}\alpha^3
 -{ 6477 \over 32}\Lambda^{15}\alpha^4
 -\cdots
 \cr
 W^{{\rm DV},Sp(8)}_{\rm traceful}(\Lambda,\alpha) &=
 5\Lambda^3
 -{5}\Lambda^6\alpha
 -{13}\Lambda^9\alpha^2
 -{65}\Lambda^{12}\alpha^3
 -{2142\over 5}\Lambda^{15}\alpha^4
 -\cdots
 }
}

\subsec{Sp(N) with traceless antisymmetric matter}

Including the contribution from the disconnected propagator, the
perturbative part of the glueball superpotential for $Sp(N)$ with
traceless antisymmetric matter is
\eqn\bo{\eqalign{
W_{\rm traceless}^{\rm pert}(S,\alpha)
=&
\left(-1+{4\over N}\right)\alpha S^2
+\left(-{1\over 3}-{8\over N}+{160\over 3N^2}\right)\alpha^2 S^3\cr
&+\left(-{1\over 3}-{12\over N} -{256 \over 3N^2}+{3584 \over 3N^3}\right)\alpha^3 S^4+\cdots
}}
The presence of many disconnected diagrams makes this case more
complicated than the traceful case, and we have correspondingly
worked to one lower order than in \bm.

 Adding the
Veneziano--Yankielowicz superpotential and integrating out the
glueball superfield, we obtain
\eqn\boa{\eqalign{
 W_{\rm traceless}^{\rm DV}(\Lambda,m,g)
 =&
 (N/2+1)\Lambda^3
 \Bigg[
 1
 -\frac{2(N-4)}{N(N+2)}\Lambda^3\alpha
 -\frac{2(N^3+14N^2-16N-512)}{3N^2(N+2)^2}\Lambda^6\alpha^2\cr
 &-\frac{2(N+8)^2(N^3+12N^2-52N-528)}{3N^3(N+2)^3}\Lambda^9\alpha^3
 -\cdots
 \Bigg]
}}
This yields
\eqn\bq{\eqalign{
 W^{{\rm DV},Sp(4)}_{\rm traceless}(\Lambda,\alpha)
 &=
 3\Lambda^3
 +{}\Lambda^9\alpha^2
 +{10}\Lambda^{12}\alpha^3
 +\cdots
 \cr
 W^{{\rm DV},Sp(6)}_{\rm traceless}(\Lambda,\alpha)
 &=
 4\Lambda^3
 -{1\over 3}\Lambda^6\alpha
 -{7\over 54}\Lambda^9\alpha^2
 +{49 \over 54}\Lambda^{12}\alpha^3
 +\cdots
 \cr
 W^{{\rm DV},Sp(8)}_{\rm traceless}(\Lambda,\alpha)
 &=
 5\Lambda^3
 -{1\over 2}\Lambda^6\alpha
 -{2\over 5}\Lambda^9\alpha^2
 -{14\over 25}\Lambda^{12}\alpha^3
 -\cdots
 }
}

\newsec{Gauge theory example: Sp(N) with antisymmetric matter}

Dynamically generated superpotentials can be determined for ${\cal
N}=1$ theories with gauge group $Sp(N)$ and a chiral superfield
$A_{ij}$ in the antisymmetric tensor representation. The general
procedure was given in \ChoBI, and is reviewed in  Appendix C.
Since these superpotentials cannot be obtained by the integrating
in procedure of \IntriligatorUK, they are more difficult to
establish, and the results are correspondingly more involved, than
more familiar examples.
  A separate computation is required for each $N$, and
the results display no obvious pattern in $N$.  $N=4$ is a simple
special case (since $Sp(4) \approx SO(5)$ and $A_{ij} \approx $
vector); $N=6$ was worked out in  \ChoBI, and in Appendix C we
extend this to $Sp(8)$ (the second paper listed in \ChoBI gives the result
for $Sp(8)$ with some additional fundamentals, which need to be integrated
out for our purposes).   In this section we state the results,
and integrate out $A_{ij}$ to obtain formulas that we can compare
with the DV approach.

The moduli space of the classical theory is parameterized by the
 gauge invariant operators
\eqn\ea{
 O_n=\Tr[(A J)^n],\qquad n = 1, 2, \ldots , N/2
}
the upper bound coming from the characteristic equation of the
matrix $AJ$.

${}$ From the gauge theory point of view, it is natural to demand
tracelessness, and this will be denoted by a tilde:  $\Tr [\At J]=0$,
\eqn\eab{
 \Ot_n=\Tr[(\At J)^n],\qquad n =  2, \ldots , N/2~.
}
In comparing with the DV approach, we will consider both the traceless
and traceful cases.

\subsec{Traceless case}

The $Sp(4)$ and $Sp(6)$ dynamical superpotentials for these
fields are \ChoBI:
\eqn\eb{
 W_{\rm dyn}^{Sp(4)}
={2\Lambda_0{}^4 \over \Ot_2^{1/2}},
}
\eqn\ec{
 W_{\rm dyn}^{Sp(6)}
={4\Lambda_0{}^5 \over \Ot_2 [(\sqrt{R}+\sqrt{R+1})^{2/3}+
 (\sqrt{R}+\sqrt{R+1})^{-2/3}-1]},
}
with $R= -12 \Ot_3^{~2}/\Ot_2^{~3}$.

Also, as derived in Appendix C, the $Sp(8)$ superpotential is
\eqn\ed{
 W_{\rm dyn}^{Sp(8)}
 = \frac{6\sqrt{2}\,\Lambda_0^6}{\Ot_2^{3/2}}
 \bigl[
 -36 R_4
 +144 b^2 R_4
 +288 c R_4
 +8 R_3^2
 +192 bc R_3
 +1152b^2c^2
 -36 b^2
 -72 c
 +9
 \bigr]^{-1},
}
where $R_3\equiv \Ot_3/\Ot_2{}^{3/2}$, $R_4\equiv \Ot_4/\Ot_2{}^2$, and
$b$ and $c$ are determined by
\eqn\ef{\eqalign{
 12R_4  +16bR_3 -192b^2c +24b^2 +96c^2 -3  &=0,\cr
 12bR_4 +8b^2R_3 +8R_3c -96bc^2  +24bc -3b  &=0.
}}
We choose the root which gives $R_3=0$ as the solution of the
$F$-flatness condition.

Now let us integrate out the antisymmetric matter.
We add the tree level superpotential
\eqn\eg{
 W_{\rm tree}=\frac{m}{2}\Ot_2+\frac{g}{3}\Ot_3
}
to the dynamical part, solve the $F$-flatness equations, and substitute
back in.   We do this perturbatively in $g$, and obtain
\eqn\eh{\eqalign{
 W^{{\rm gt}, Sp(4)}_{\rm traceless}
 &=
 3\Lambda^3,
 \cr
 W^{{\rm gt}, Sp(6)}_{\rm traceless}
 &=
 4\Lambda^3
 -{1 \over 3}\Lambda^6\alpha
 -{7 \over 54}\Lambda^9\alpha^2
 -{5 \over 54}\Lambda^{12}\alpha^3
 -{221 \over 2592}\Lambda^{15}\alpha^4
 -\cdots,
 \cr
 W^{{\rm gt}, Sp(8)}_{\rm traceless}
 &=
 5\Lambda^3
 -{1 \over 2}\Lambda^6\alpha
 -{2 \over 5}\Lambda^9\alpha^2
 -{14 \over 25}\Lambda^{12}\alpha^3
 -{}\Lambda^{15}\alpha^4
 -\cdots,
}}
where $\alpha$ is defined in \bmbb, and the low-energy scales
are defined from the usual matching conditions as
\eqn\eha{\eqalign{ 
(\Lambda^3)^{\frac{N}{2}+1}=\left(\frac{m}{2}\right)^{\frac{N}{2}-1}
\Lambda_0^{N+4}.
}}

\subsec{Traceful case}

For $Sp(N)$ theory with a traceful antisymmetric tensor $A_{ij}$,
we separate out the trace part as
\eqn\ep{
A_{ij}=\At_{ij}-\frac{1}{N}J_{ij}\phi,\qquad
\Tr[\At J]=0,\qquad
\Tr[AJ]=\phi.
}
$\Ot_n$ are related to their traceful counterparts $O_n\equiv\Tr[(AJ)^n]$ by
\eqn\eq{
O_2=\Ot_2+{\phi^2 \over N},\qquad
O_3=\Ot_3
+{3 \over N}\Ot_2 \phi
+{1 \over N^2} \phi^3.
}
The dynamical superpotential of this traceful theory is the same as the
traceless theory, since $\phi$ has its own $U(1)_{\phi}$ charge and
hence cannot enter in $W_{\rm dyn}$.

Integrating out $\At_{ij}$ and $\phi$ in the presence of the tree level
superpotential
\eqn\er{
 W_{\rm tree}=\frac{m}{2}O_2+\frac{g}{3}O_3,
}
we obtain
\eqn\es{\eqalign{
 W^{{\rm gt}, Sp(4)}_{\rm traceful}
 &=
 3\Lambda^3
 -{}\Lambda^6\alpha
 -{2}\Lambda^9\alpha^2
 -{187 \over 27}\Lambda^{12}\alpha^3
 -{2470 \over 27}\Lambda^{15}\alpha^4
 -\cdots,
 \cr
 W^{{\rm gt}, Sp(6)}_{\rm traceful}
 &=
 4\Lambda^3
 -{3}\Lambda^6\alpha
 -{47 \over 6}\Lambda^9\alpha^2
 -{75 \over 2}\Lambda^{12}\alpha^3
 -{7437 \over 32}\Lambda^{15}\alpha^4
 -\cdots,
 \cr
 W^{{\rm gt}, Sp(8)}_{\rm traceful}
 &=
 5\Lambda^3
 -{5}\Lambda^6\alpha
 -{13}\Lambda^9\alpha^2
 -{65}\Lambda^{12}\alpha^3
 -{2147 \over 5}\Lambda^{15}\alpha^4
 -\cdots,
}}

\newsec{Comparison and Discussion}

According to the general conjecture, we are supposed to compare \bq\
with \eh, and \bn\ with \es.    We write $\bigtriangleup W\equiv
W^{\rm DV}-W^{\rm gt}$, and find
\eqn\ek{\eqalign{
\bigtriangleup W_{\rm traceless}^{Sp(4)}
&= 
0 \cdot \L^6 \a
+\Lambda^9 \alpha^2
+10\Lambda^{12} \alpha^3
+\cdots,
\cr
\bigtriangleup W_{\rm traceless}^{Sp(6)} 
&=
0 \cdot \L^6 \a
+  0 \cdot \L^9 \a^2
+\Lambda^{12}\alpha^3
+\cdots,
\cr
\bigtriangleup W_{\rm traceless}^{Sp(8)} 
&=
0 \cdot \L^6 \a
+  0 \cdot \L^9 \a^2
+ 0 \cdot \L^{12} \a^3
+{\cal O}(\Lambda^{15 }\alpha^4).
}}
and
\eqn\eka{\eqalign{
\bigtriangleup W_{\rm traceful}^{Sp(4)}
&= 
0 \cdot \L^6 \a
+\Lambda^9 \alpha^2
+20\Lambda^{12} \alpha^3
+{32615\over 81}\Lambda^{15} \alpha^4
+\cdots,
\cr
\bigtriangleup W_{\rm traceful}^{Sp(6)} 
&=
0 \cdot \L^6 \a
+  0 \cdot \L^9 \a^2
+\Lambda^{12}\alpha^3
+30\Lambda^{15}\alpha^4
+\cdots,
\cr
\bigtriangleup W_{\rm traceful}^{Sp(8)} 
&=0 \cdot \L^6 \a
+ 0 \cdot \L^9 \a^2
+ 0 \cdot \L^{12} \a^3 
+\Lambda^{15 }\alpha^4+\cdots.
}}

We have indicated the terms that cancelled nontrivially by including
them with a coefficient of  zero. From these examples, 
we see that a disagreement sets in at order
$( \L^3)^h\a^{h-1}$, where $h=N/2+1$ is the dual Coxeter number.
We also observe that the coefficient of the disagreement at
this order is unity.  We now discuss the implications of this result.

First, it is very unlikely that the discrepancy is due to a 
computational error, such as forgetting to include a diagram.  This is
apparent from the fact that the mismatch arises at a different order
in perturbation theory  for different rank gauge groups.  So adding a new
contribution to the $Sp(4)$ result at order $\L^9 \a^2$, say, would
generically destroy the agreement for $Sp(6)$ and $Sp(8)$ at this order.
Instead, it is much more likely that our results indicate a breakdown
of the underlying approach.

Let us return to the two basic elements of the DV conjecture.  The first
part asserts that the perturbative part of the glueball superpotential
can be computed from matrix integrals, and the second part assumes that
nonperturbative effects are captured by adding
the Veneziano-Yankielowicz superpotential.  We have proven the 
perturbative part of the conjecture for the relevant matter fields,
but there are  subtleties which we have so far avoided but now 
must discuss.

In our perturbative computations we inserted no more than two $\CW_\a$'s
on any index loop, since we were interested in a superpotential for
$S \sim \Tr \CW^2$, and not in operators such as $\Tr ~\CW^{2n}$, $n>1$.  
However, for a given gauge group, it may be possible to use Lie algebra 
identities to express such ``unwanted'' operators in terms of other operators,
including $S$.  Should we then include these new $S$ terms along with our
previous results? 

This question might make one  suspicious of 
the usual procedure, especially in our case given
that the discrepancy sets in at  order $S^h$, which is when we begin
to find nontrivial relations involving $S$ due to its underlying
structure as a product of the fermionic field $\CW_\alpha$.  For example,
for $Sp(4)$ there are relations such as
\eqn\fa{
\Tr[(\CW^2)^3]=
\frac{3}{4}\Tr[\CW^2]\Tr[(\CW^2)^2]
-\frac{1}{8}(\Tr[\CW^2])^3.}
So a naive guess is that the discrepancies can be accounted for if we keep
all contributions coming from more than two $\CW_\alpha$'s on an index
loop, and re-express the traces of the form $\Tr(\CW_\alpha)^{2n}$
($n\ge h$) in favor of $S$ using relations like \fa, setting all traces
to zero that are not re-expressible in terms of $S$.  
Such considerations are indeed necessary in order to avoid getting
unexpected results in certain cases, e.g.\ antisymmetric matter for
$Sp(2)$.   Such a matter field is uncharged, and so one would expect
(but see the discussion below) it to 
contribute a vanishing result for the glueball superpotential, but this
is  seen only if we compute all the trace structures.  We should 
emphasize  that if we keep 
all of these contributions then perturbation theory will not reduce
to matrix integrals since the Schwinger parameter dependence will not
cancel; nevertheless we can try this procedure and see what we get.  

In order to check whether the above guess could be correct, 
we took a $\Phi^{2p}$
interaction and evaluated the perturbative superpotential explicitly,
keeping all the traces.  For this interaction, a discrepancy arises at the
first order if we take $p>N$.  Specifically, we considered a $\Phi^{6}$
interaction in $Sp(4)$ with antisymmetric matter. 
After a tedious calculation, we have found that 
this  does {\it not} account for the discrepancy.  In retrospect this is 
not really surprising,  for two reasons. First,
relations like \fa\ are corrected
nonperturbatively.  Second it is not obviously correct to set to zero
terms like the one appearing in the middle of \fa.  
One argument that we need not include such  operators
is that they have vanishing expectation values,
so in the equations of motion they can be set to zero, leaving an
equation of motion for $S$.  However, this does not
necessarily justify setting these operators to zero in the Lagrangian
before deriving the equations of motion for $S$.  
For example, when we encounter an expression like
\fa, we might have to replace it as
\eqn\faa{
\Tr[(\CW^2)^3]\to
{\cal P}^{(3)}(S,\Lambda,\alpha)
}
where ${\cal P}^{(3)}$ is some polynomial in $S$ of degree three that
vanishes on shell, and include it in the effective superpotential.



Other nonperturbative subtleties may also play a role.\foot{We thank
Cumrun Vafa for bringing these issues to our attention.}
Even if the basic procedure is correct, discrepancies might seem to
arise due to a redefinition of couplings and operators involved
in translating  between gauge theory and matrix model expressions.  
It was argued in \DoreyTJ\ that this is what happens in the 
${\cal N}=1^*$ theory.  In that theory there were some constraints
which could be imposed on possible operator redefinitions, but it is
not clear whether this can also be done in our case.  To check this
one needs to consider all possible operator mixings, including mixing
of single-trace  and multi-trace  operators.  

It is also possible that the matrix model computation 
corresponds to a different gauge theory than the one we have been comparing
with.  In particular, starting from a string theory construction it is 
possible that there are some exotic nonperturbative 
effects which survive the field theory
limit.  In this case, matrix model results should be compared against
theories in a different universality class than ordinary gauge theories. 
From this point of view it may be possible to justify nontrivial results
for seemingly trivial theories, {\it e.g.} $Sp(2)$ with traceless
anti-symmetric matter.  It would therefore be very instructive to
find a string theory realization of our theories.

To summarize, 
our results indicate that for generic theories the simplest form of the
DV conjecture is valid  up to $(h-1)$ loops.   
On the other hand, the fact that
our discrepancies arise in a very simple fashion --- always with a
coefficient of unity --- suggests that perhaps there exists a way 
of modifying the DV recipe to enable us to go to $h$ loops and 
beyond.
Clearly, it is important to resolve these issues in order to determine
the range of validity of the DV approach.  One might  hope that
the approach will be useful for any ${\cal N}=1$ theory admitting a
large $N$ expansion.  Our $Sp(N)$ theories are certainly in this class,
and so provide an important challenge.

\bigskip\medskip\noindent
{\bf Acknowledgements:} 
We thank Iosif Bena, Eric D'Hoker, Anton Ryzhov, and Cumrun Vafa for
helpful discussions. This work was supported by NSF grant PHY-0099590.

\newsec{Appendix A: diagrammatics for traceless matter field}

In this appendix, we sketch the diagrammatics for evaluating the
perturbative glueball superpotential, focusing on the case with a
traceless tensor.  To be specific, we consider the cubic interaction
below:
\eqn\yaa{
e^{-W^{\rm pert}(S)}=\int \CD\Phi\, e^{
-\int d^4x\, d^2\theta\,\Tr\left[
-\frac{1}{2}\Phi(\p^2-i\CW^\alpha D_\alpha)\Phi
+\frac{m}{2}\Phi^2
+\frac{g}{3}\Phi^3
\right]
}.
}
Namely, we consider $SO$ with tracelss symmetric matter, or
$Sp$ with traceless antisymmetric matter.

%
%
\fig{Diagrams for traceless tensor matter field}{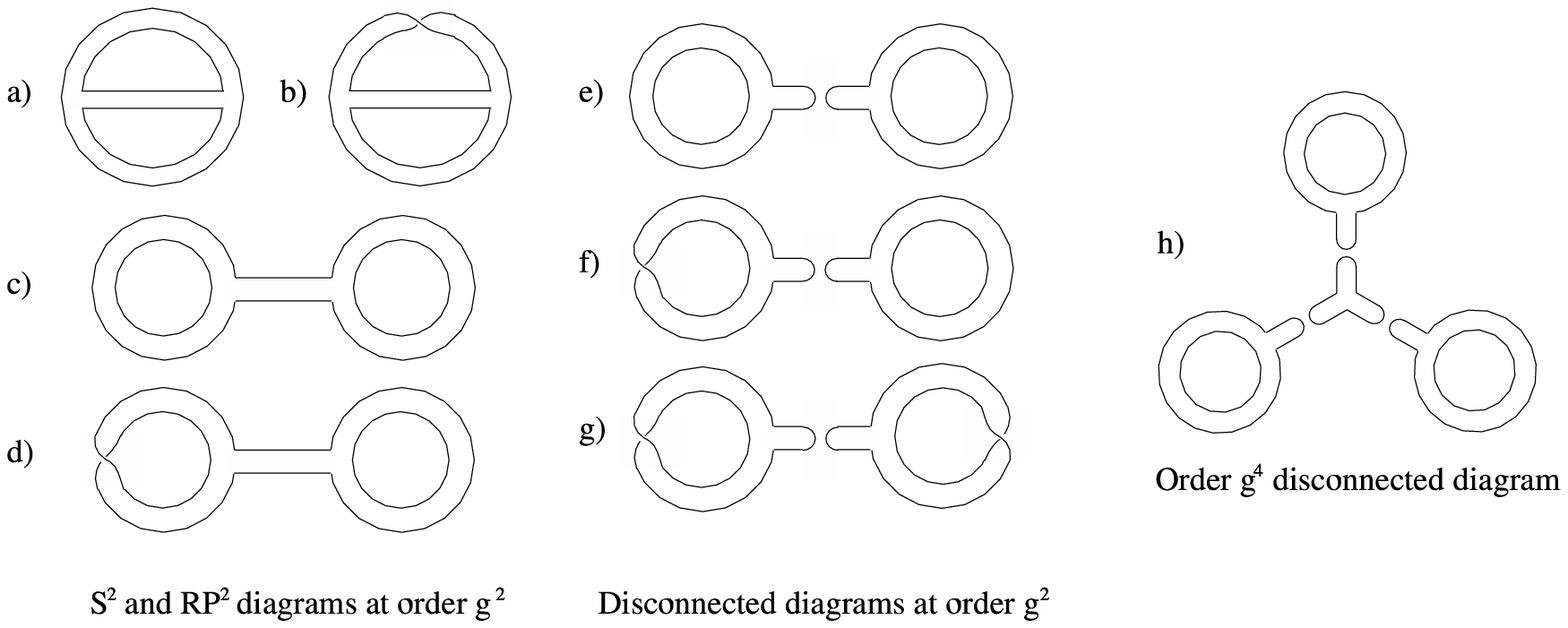}{14cm}
\figlabel{\diagfig}

At order $g^2$, there are four $S^2$ and $RP^2$ diagrams without
disconnected propagators that contribute, as shown in Fig.\
\diagfig\ a)--d).  These can be evaluated  by combinatorics.  For
a) and b) there are $6$ ways to contract legs.  For c) and d), on
the other hand, there are $3^2$ ways to contract legs and 2
choices for the middle propagator (untwisted or twisted).  Since
there are two loop momenta, the glueball $S\sim \CW^\alpha
\CW_\alpha$ should be inserted in two index loops, and we may
insert only up to one glueball on each index loop. For $S^2$
graphs a) and c), there are 3 ways to do so, and a trace over the
remaining index loop contributes $N$.  For $RP^2$ graphs b) and
d), there is only one way to insert the glueball, but the
fermionic determinant gives an extra factor $(\pm 4)$.  The sign
depends on the matter field under consideration {see section 2.)
Finally, for b) and d) there are respectively 3 and 2 ways to
choose which propagator to twist. Therefore, the contributions are
\eqn\ya{
f_{\rm a}=6\cdot 3\cdot N S^2,\quad
f_{\rm b}=6 \cdot 3\cdot (\pm 4) S^2,\quad
f_{\rm c}=3^2\cdot 2\cdot 3\cdot N S^2,\quad
f_{\rm d}=3^2\cdot 2\cdot 2 \cdot (\pm 4) S^2.
}
Including factors coming from propagators and coefficients from Taylor
expansion, we obtain
\eqn\yab{
W^{(2)}_{\rm conn}=-{g^2 \over (2m)^3}{1 \over 2! \,3^2}(f_{\rm a}+f_{\rm b}+f_{\rm c}+f_{\rm d})
= (-N\mp 3)\alpha S^2.
}
where we defined  $\alpha =\frac{g^2}{2m^3}$ as before. This
reproduces the first term of the traceful result \bm.

For a traceless  tensor, there are three additional
diagrams e), f) and g),  with disconnected propagators that give
nonvanishing contributions.

These can be evaluated similarly to the connected ones.  First, there
are factors common to all three graphs; $(-2/N)$ from the disconnected
propagator, and $3^2=9$ from the ways to contract legs.  In addition,
the particular graphs have the additional factors;\ \ e): $(2N)^2$ from 
the ways of inserting a glueball in one of two index loops in each
$S^2$ component, and the trace on the remaining index loop.  f): $(\pm 4)$ from
the fermionic determinant of the $RP^2$ component, and $2N$ from the
glueball insertion into the $S^2$ component.  Also, there is the same
contribution from the $S^2\times RP^2$ graph.  g): $(\pm 4)^2$ from two
$RP^2$ components.  Altogether we obtain
\eqn\yc{
W^{(2)}_{\rm disconn}
=
-{g^2 \over (2m)^3}{1 \over 2! \,3^2}
\left(-\frac{2}{N}\right)\cdot 9\cdot
[(2N)^2+2\cdot 2N\cdot(\pm 4)+(\pm 4)^2]S^2
={(2N\pm 4)^2\over 4N}\alpha S^2.
}
Summing the connected and disconnected contributions, we obtain
\eqn\yc{
W^{(2)}_{\rm conn}+
W^{(2)}_{\rm disconn}
=
\left(\pm 1+\frac{4}{N}\right)\alpha S^2
}
which is the first term of the traceless result \bo.

Higher order diagrams can be worked out in  much the same way,
although the number of diagrams increases rapidly.  For the
disconnected diagrams, we only have to consider the diagrams which are
one-particle-reducible with respect to the disconnected propagator.
Therefore, we basically  just splice lower order diagrams with the
disconnected propagator.  The contribution is just the product of the
contributions from the lower order pieces, multiplied by the ways to
insert the disconnected propagator into them, and by $(-2/N)^n$ from
the disconnected propagator itself.  However, note that one should also
consider diagrams such as h) of Fig.\ \diagfig.  In this case, the
central $D_2$ piece contributes $N$ from its index loop.

\newsec{Appendix B:  summary of results from perturbation theory}

In this appendix we state our results for $W^{\rm pert}(S,\a)$, the
perturbative contributions to the glueball superpotential.  These correspond
to evaluating certain diagrams in the matrix model.  We consider
cubic interactions only,
\eqn\xa{ W_{\rm tree}=\frac{m}{2}\Tr\,\Phi^2 +\frac{g}{3} \Tr\,\Phi^3,}
which means that we will not consider the case of $SO/Sp$ with
adjoint matter. In any event, it is not necessary to compute the
perturbative superpotential for the latter cases, since closed
form expressions for even power interactions are already known
\refs{\ItaKX,\AshokBI,\JanikNZ}.  The case of $U(N)$ with adjoint
matter is also well known \BrezinSV, but for convenience we include it in
the list below.

For traceful matter fields, instead of evaluating individual
Feynman diagrams, there exists a much simpler method for computing
which we have used to obtain the results below.  We can simply
compute the matrix model free energy by computer for certain low
values of $N$.  Since $S^2$ and $RP^2$ diagrams scale as $N^{L+1}$
and $N^L$ at $L$ loops in perturbation theory, we can easily read
off the $S^2$ and $RP^2$ contributions to any desired order.  For
traceless fields things are not so simple, since the $N$
dependence becomes more complicated, and certain diagrams must be
discarded (as discussed in Section 2.)

We define
\eqn\xaa{\a = \left\{\eqalign{g^2/m^3 & \quad\quad U(N) \cr
 g^2 /2m^3 &\quad\quad SO/Sp(N)}\right.}

\subsec{U(N) with adjoint matter}
\vskip-.5cm
\eqn\xe{\eqalign{W^{\rm pert}(S,\alpha) & = N {\p F_{\chi=2} \over \p S},
\quad\quad  F_{\chi=2}= -{S^2 \over 2} \sum_{k=1}^\infty
{ (8 \a S)^k \over (k+2)!}{\Gamma\left({3k \over 2}\right) \over
\Gamma\left({k \over 2}+1 \right)} \cr W^{\rm pert}(S,\alpha)& = -2N\a S^2
-{32 \over 3}N \a^2 S^3 - {280 \over 3} N \a^3 S^4 - 1024 N \a^4 S^5 - \cdots}}

\subsec{SU(N) with adjoint matter}
\vskip-.5cm
\eqn\xf{
W^{\rm pert}(S,\alpha) =0\cdot \alpha S^2+0\cdot \alpha^2 S^3+
0\cdot \alpha^3 S^4 + \cdots  }

\subsec{SO(N) with traceful symmetric matter}
\vskip-.4cm
\eqn\xd{\eqalign{
W^{\rm pert}(S,\alpha)
&=
-\left(N+3\right)\alpha S^2
-\left({16\over 3}N+{59\over 3}\right)\alpha^2 S^3\cr
&-\left({140\over 3}N+197\right)\alpha^3 S^4
-\left({512}N+{4775\over 2}\right)\alpha^4 S^5
-\cdots}
}

\subsec{SO(N) with traceless symmetric matter}
\eqn\xm{\eqalign{
W^{\rm pert}(S,\alpha)
=&
\left(1+{4\over N}\right)\alpha S^2
+\left({1\over 3}-{8\over N}-{160\over 3N^2}\right)\alpha^2 S^3\cr
&+\left({1\over 3}-{12\over N} +{256 \over 3N^2}
+{3584 \over 3N^3}\right)\alpha^3 S^4+\cdots
}}

\subsec{Sp(N) with traceful antisymmetric matter}
\vskip-.5cm
\eqn\xb{\eqalign{
W^{\rm pert}(S,\alpha)
&=
\left(-N+3\right)\alpha S^2
+\left(-{16\over 3}N+{59\over 3}\right)\alpha^2 S^3\cr
&+\left(-{140\over 3}N+197\right)\alpha^3 S^4
+\left(-{512}N+{4775\over 2}\right)\alpha^4 S^5
+\cdots}
}

\subsec{Sp(N) with traceless antisymmetric matter}
\vskip-.5cm
\eqn\xc{\eqalign{
W^{\rm pert}(S,\alpha)
=&
\left(-1+{4\over N}\right)\alpha S^2
+\left(-{1\over 3}-{8\over N}+{160\over 3N^2}\right)\alpha^2 S^3\cr
&+\left(-{1\over 3}-{12\over N} -{256 \over 3N^2}
+{3584 \over 3N^3}\right)\alpha^3 S^4+\cdots
}}

These results exhibit some remarkable cancellations.  We find a
vanishing result for $SU(N)$ with adjoint matter, and a cancellation of
the terms linear in $N$ for $Sp(N)$ with traceless antisymmetric matter.
In both cases the cancellation seems to involve all the diagrams at a
given order; for example, in Fig.\ 5, the $\CO(N)$ contribution from e)
cancels the $\CO(N)$ contributions from both a) and c).  Also, the
cancellation does not seem to be special to cubic interactions; we have
checked at leading order that the cancellation occurs for $\Phi^5$ and $\Phi^7$
interactions, even when both are present
at the same time.  Therefore the cancellation probably occurs for any
tree level superpotential with odd power terms only.  We do not have a
proof of cancellation beyond the order indicated; it would be nice to
provide one and to better understand the significance of this fact.

\newsec{Appendix C: gauge theory results}

In \ChoBI, a systematic method for determining the dynamical superpotential of
the $Sp(N)$ gauge theory with a traceless antisymmetric matter
$\At_{ab}$ was proposed.  In this appendix, we briefly review the
strategy, focusing on the $Sp(8)$ case.

First, we add to the theory $2N_F$ fundamentals $Q_i$.  The
moduli space of this enlarged $(N_{\At},N_F)$ theory is parameterized by

\eqn\xa{
 \Ot_n=\Tr[(\At J)^n],\qquad n=2,3,\cdots, N/2
}
as well as the antisymmetric matrices
\eqn\xb{
 M_{ij}=Q_i^T J Q_j,\
 N_{ij}=Q_i^T J\At J Q_j,\
 P_{ij}=Q_i^T J(\At J)^2 Q_j,\
 \cdots,\
 R_{ij}=Q_i^T J(\At J)^{k-1} Q_j.
}
The basic observation is that for $N_F=3$,  symmetry and
holomorphy considerations restrict the dynamical superpotential to
be of the form
\eqn\xc{
 W_{(1,3)}^{\rm dyn}
 = \frac{{\rm Some\ polynomial\ in\ }\Ot_n, M_{ij}, \dots, R_{ij}}{\Lambda_{(1,3)}^{b_0}},
}
where $b_0=N-N_F+4=N+1$ and the subscript $(1,3)$ denotes the matter
content $(N_{\At},N_F)$.  The polynomial must of course respect
the various symmetries of the theory.    More significantly, the
$F$-flatness equations following from $W_{(1,3)}^{\rm dyn}$ can be written in a
$\L$ independent form.   By setting $\L=0$, one sees that the equations must
reduce to the classical constraints which follow upon expressing
the gauge invariant field $\Ot_n$, $M_{ij}$, $\dots$, $R_{ij}$  in terms
of their constituents $Q_i$ and $\At$.
 Quantum corrections to these classical constraints are forbidden
by symmetry and holomorphy. These requirements fix $W_{(1,3)}^{\rm
dyn}$ up to an overall normalization.  Once we have obtained
$W_{(1,3)}^{\rm dyn}$, we can derive the desired $W_{(1,0)}^{\rm
dyn}$ by giving mass to $Q_i$ and integrating them out.

For $Sp(8)$, the above procedure uniquely determines the
superpotential to be (this result appears in the second paper of \ChoBI)
\eqn\xd{\eqalign{
 W_{(1,3)}^{\rm dyn}
 =&\frac{1}{\Lambda^9_{(1,3)}}
 \Big[
 1152(PPP)+6912(RPN)+3456(RRM) -864 \Ot_2 (PNN)
 \cr
 &-1728 \Ot_2(RNM)+108 \Ot_2^2(NNM) -108 \Ot_2^2(PMM)+9 \Ot_2^3 (MMM)\cr
 &+192 \Ot_3(NNN)-576 \Ot_3 (RMM) + 144 \Ot_2 \Ot_3 (NMM) +32
 \Ot_3^2(MMM)\cr
 &+432 \Ot_4 (NNM)+432 \Ot_4(PMM)-36 \Ot_2 \Ot_4 (MMM) \Big],
}}
up to normalization, where $(ABC)\equiv
\epsilon^{ijklmn}A_{ij}B_{kl}C_{mn}$.  Now that we have obtained
$W_{(1,3)}^{\rm dyn}$, we can integrate out $Q_i$ by adding a mass term
\eqn\xe{
 W_{(1,3)}^{\rm dyn}\to W_{(1,3)}^{\rm dyn}+\frac{\mu^{ij}}{2} M_{ij}.
}
When solving the $F$-flatness condition, we can assume that $M_{ij},
N_{ij}, P_{ij}, R_{ij}\propto (\mu^{-1})_{ij}$ since $\mu^{ij}$ is the
only quantity they can depend on.  Plugging back in, we
obtain the $Sp(8)$ superpotential \ed\ and \ef.  The same procedure
leads to the superpotential \eb\ and \ec\ for $Sp(4)$ and $Sp(6)$,
respectively \ChoBI.

\listrefs

\end